\documentclass[10pt,preprint2]{aastex}

\newcommand{\CaII}{\ion{Ca}{2} }

\shorttitle{The Puzzlingly Large \CaII Triplet}
\shortauthors{Michielsen {\em et al.}}

\begin{document}

\title{The puzzlingly large \CaII triplet absorption in dwarf
       elliptical galaxies\altaffilmark{1}}

\author{D. Michielsen, S. De Rijcke\altaffilmark{2}, H. Dejonghe,}
\affil{Sterrenkundig Observatorium, Ghent University, Krijgslaan
281, S9, B-9000 Gent, Belgium}
\email{dolf.michielsen@UGent.be}
\email{sven.derijcke@UGent.be} 
\email{herwig.dejonghe@UGent.be}
\author{W.~W. Zeilinger,}
\affil{Institut f\"ur Astronomie, Universit\"at Wien,
T\"urkenschanzstra{\ss}e 17, A-1180 Wien, Austria}
\email{zeilinger@astro.univie.ac.at}
\author{G.~K.~T. Hau}
\affil{ESO, Alonso de Cordova 3107, Santiago, Chile}
\email{ghau@eso.org} 

\altaffiltext{1}{Based on observations collected at the European
    Southern Observatory, Paranal, Chile (ESO Large Program
    165.N~0115)} 
\altaffiltext{2}{Research Postdoctoral Fellow of the Fund for
    Scientific Research - Flanders (Belgium)(F.W.O)}

\begin{abstract}
We present central CaT, PaT, and CaT* indices for a sample of fifteen dwarf
elliptical galaxies (dEs). Twelve of these have CaT*$\sim 7$~{\AA} and
extend the negative correlation between the CaT* index and central
velocity dispersion $\sigma$, which was derived for bright
ellipticals (Es), down to $20<\sigma<55$~km/s. For five dEs we have
independent age and metallicity estimates. Four of these have
CaT*$\sim 7$~{\AA}, much higher than expected from their low
metallicities ($-1.5<$ [Z/H] $<-0.5$). The observed anti-correlation
of CaT* as a function of $\sigma$ or Z is in flagrant disagreement
with theory. We discuss some of the amendments that have been proposed
to bring the theoretical predictions into agreement with the observed
CaT*-values of bright Es and how they can be extended to incorporate
also the observed CaT*-values of dEs. Moreover, 3 dEs in our sample
have CaT*$\sim 5$~{\AA}, as would be expected for metal-poor stellar
systems. Any theory for dE evolution will have to be able to explain
the co-existence of low-CaT* and high-CaT* dEs at a given mean
metallicity. This could be the first direct evidence that the dE
population is not homogeneous, and that different evolutionary paths
led to morphologically and kinematically similar but chemically
distinct objects.
\end{abstract}

\section{Introduction:~the low CaT-value of high-mass galaxies}

The \CaII triplet (8498, 8542, 8662~{\AA}) is a prominent
absorption-line feature in the near-infrared spectrum of cool
stars. Theoretical and empirical population synthesis modelling of the
\CaII triplet showed that it is a good tracer of the metallicity of
stellar systems \citep{id97,gar98}. This is indeed observed to be the
case for Galactic globular clusters \citep{az88,rut97}. However, in
early type galaxies only a small spread of \CaII strengths was
measured \citep{coh79,ba87,ter90}.

Recently, \citet{cen01} defined new line-strength indices for the
strength of the \CaII triplet (CaT) and for the combined strength of
the P12, P14, and P17 H Paschen lines (PaT). The \CaII index
corrected for the contamination by the Paschen P13, P15, and P16 lines
is denoted by CaT* (CaT*=CaT$-0.93 \times$PaT). These authors compiled
a large library of stellar spectra and produced fitting functions
\citep{cen02} that can be employed to predict index-values for
single-age, single-metallicity stellar populations (SSPs) using
population synthesis models \citep{vaz03}. These models predict that
for low metallicities, CaT* should be sensitive to metallicity but
virtually independent of age.

However, \citet{sag02} (SAG) and \citet{cen03} (CEN) present CaT*
indices for bright ellipticals and show that CaT* and central velocity
dispersion $\sigma$ anticorrelate. \citet{fal03} (FAL) arrive at the
same conclusion based on a sample of bulges of spirals. On a linear
$\sigma$-scale, all samples yield essentially the same slope and
zeropoint. These results provide the first evidence for an
anti-correlation between a metal-line index and $\sigma$ while other
metallicity indicators such as Mg$_2$ increase with $\sigma$
\citep{ter90}. Moreover, the CaT* values of the Es in the SAG-sample
scatter around 6.9~{\AA}, about 0.5~{\AA} lower than expected by SSP
models, given their ages and metallicities (determined independently
from Lick indices). CEN do not give independent metallicities for
their sample but find that the measured CaT* values, with the
exception of those of a few low-mass Es, are lower than any model
prediction using a Salpeter IMF.

\section{Observations and Data Reduction} \label{par_ob}

We collected long-slit spectra of 15 dEs in the Fornax Cluster and in
the NGC5044, NGC5898, and Antlia Groups in the wavelength region
$\lambda\lambda 7900-9300$~{\AA}. These observations were carried out
at the ESO-VLT with FORS2 during 5 observing runs in 2001 and 2002. We used
the FORS2 grism GRIS\_1028z+29, which, with a 0.7$''$ slit, yields an
instrumental broadening of $\sigma_{\rm instr}\simeq 30$~km/s. Seeing
conditions were typically 0.7-1.0$''$~FWHM. Total integration times
varied between 5 and 8 hours. We obtained spectra of late G to early M
giant stars as velocity templates. The standard data reduction
procedures were performed with ESO-MIDAS\footnote{ESO-MIDAS is
developed and maintained by the European Southern Observatory}. The
spectra for each galaxy were bias-subtracted, flatfielded, corrected
for cosmic-ray events, rebinned to a linear wavelength scale
(rectifying the emission lines of the arc spectra to an accuracy of
$\approx$1~km/s~FWHM) and co-added. After sky-subtraction, the spectra
were flux calibrated using spectrophotometric standard stars observed
in the same instrumental setup.  Since our spectral resolution is
close to that of the stellar library of \citet{cen01} and the galactic
velocity dispersions were always well below 100~km/s, no corrections
for resolution effects or Doppler broadening were necessary. This was
confirmed by comparing the values of the indices for the template
stars in common with the library. We measured the CaT, CaT*, and PaT
indices, averaged over an aperture of radius $R_e/8$ (or 1$''$ for
galaxies with $R_{\rm e} < 8\arcsec$), using the definitions given in
\citet{cen01}, and extracted kinematics out to 1-2~$R_{\rm e}$
\citep{derijcke01,derijcke03a}.

\section{Results} \label{par_re}

\begin{figure}[h]
\includegraphics[clip,scale=0.70]{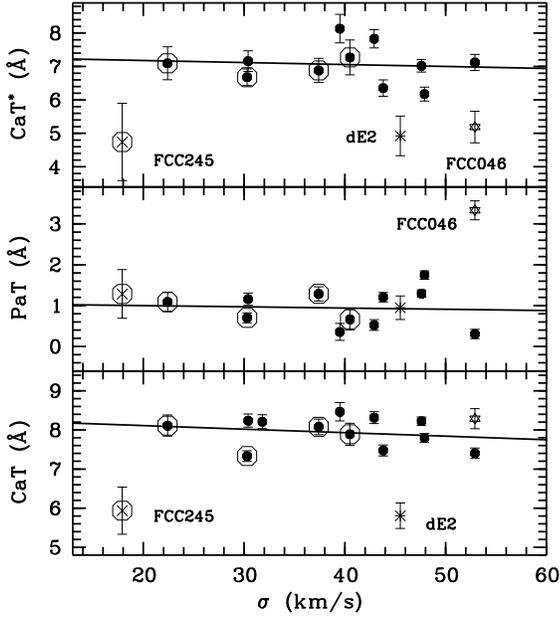}
\caption{ \footnotesize CaT, PaT and CaT* values versus central velocity
dispersion $\sigma$ for our 15 dEs with a least-squares fits to the
data, showing a mild anti-correlation. Three galaxies were excluded
from the fit: FCC245 and NGC5898\_dE2 show a very low CaT value;
FCC046 has a very high PaT absorption, due to recent starformation.
Galaxies appearing in Fig.~\ref{fig_ssp} are marked with a circle.}
\label{fig_ours}
\end{figure}

\begin{figure}[h]
\includegraphics[clip,scale=0.70]{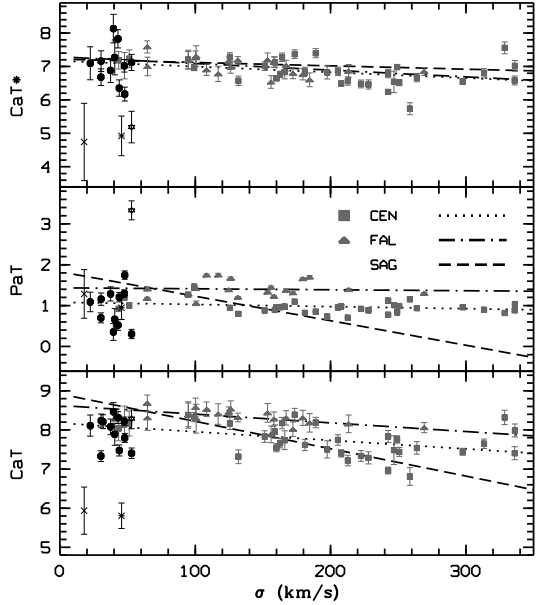}
\caption{ \footnotesize Comparison of our data with the extended relations of
SAG, CEN and FAL.
Data from CEN and FAL have been corrected for broadening and are shown together
with linear least-squares fits to their samples and the
anticorrelation found by SAG (symbols as indicated on the figure).}
\label{fig_all}
\end{figure}

The CaT, PaT and CaT* indices of our sample dEs versus central
velocity dispersion $\sigma$ are presented in Fig.~\ref{fig_ours}. The
majority of the dEs scatter around CaT*$\sim 7$~{\AA} and form the
low-$\sigma$ extension of the CaT*-$\sigma$ relation of bright Es and
bulges of spirals. Three out of fifteen dEs have a discrepantly low
CaT*$\sim 5$~{\AA} value. FCC046 has CaT=8.3$\pm$0.3~{\AA} but a very
high PaT=3.3$\pm$0.2~{\AA}, in agreement with the fact that this is an
actively star-forming dE and contains a very young stellar population
\citep{derijcke03b}. FCC245 and NGC5898\_dE2 (a dE in the NGC5898
group) have normal PaT values but very low CaT$\approx 5.5$~{\AA}.
Excluding these three galaxies, we find that CaT, PaT, and CaT* remain
nearly constant over the range $\sigma=20-55$~km/s. A detailed
analysis of the spatially resolved colors and line-strengths will be
presented in a forthcoming paper.  In Fig.~\ref{fig_all}, we present
our results together with those of SAG, CEN and FAL. CEN and FAL
broadened their spectra to 370 and 300 km/s respectively. In order to
compare their data with ours, we used the broadening corrections given
by \citet{vaz03} for an old population with solar metallicity and a
Salpeter IMF. These corrections are model-dependent, but the maximum
error introduced is $\sim 0.2$~{\AA}.  Clearly, except for the three
outliers, dEs populate the low-$\sigma$ extension of the CaT*-$\sigma$
relation of bright Es and bulges. All datasets yield approximately the
same slight CaT*-$\sigma$ anti-correlation.  SAG find PaT to
anti-correlate with $\sigma$, while the CEN data scatter around
PaT$\approx 1$~{\AA}, and FAL find PaT$\approx 1.4$~{\AA}. The same
trend is found for CaT where SAG find a stronger anti-correlation and
the Es of CEN are slightly offset with respect to the bulges of
FAL. These differences could be due in part to flux calibration
differences and to the different ways of correcting the spectra to the
system defined by the models (as discussed in FAL).

\begin{figure}[h]
\includegraphics[clip,scale=0.50]{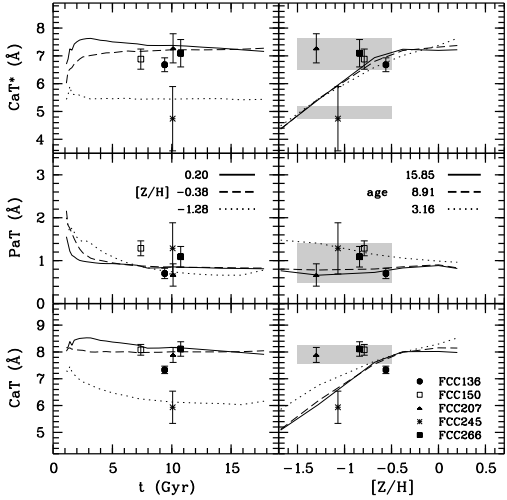}
\caption{ \footnotesize \citet{vaz03} SSP model predictions (Salpeter IMF) of CaT,
PaT and CaT* versus age (left panel) and metallicity (right panel).
The full, dashed, and dotted lines represent different metallicities
and ages as indicated on the figure. The symbols show 5 Fornax dEs of
our sample with independent age and metallicity estimations
\citep{rak01,hm94}. The shaded boxes indicate the locus of the other
dEs in our sample. The metallicity range $-1.5 \leq $[Z/H]$\leq -0.5$
is taken from \citet{rak01}, while the index ranges are 1$\sigma$
deviations from the mean. While the PaT values are consistent
with the model predictions, the near-constancy of CaT and CaT* for our
metal-poor dEs disagrees with the model predictions. Only a few dEs
have low CaT*, consistent with low metallicity (lower box in CaT*
panel).}
\label{fig_ssp}
\end{figure}

In Fig.~\ref{fig_ssp}, we show model predictions for CaT, PaT, and
CaT* as functions of age (between 1 and 18~Gyr) and metallicity
($-1.68<$ [Z/H] $<0.2$). The SSP models (Salpeter IMF) are safe for
all ages in the metallicity range $-0.7 \le$~[Z/H]~$\le 0.2$, while
for lower metallicities the age range 10-13 Gyr, which is of most
interest to us, is safe \citep{vaz03}. The predicted CaT and CaT*
values are strong functions of metallicity for [Z/H]$\le -0.4$ and
nicely reproduce the observed behavior of these indices in globular
clusters \citep{az88}, which, for all practical purposes, are genuine
SSPs. However, the bright ellipticals in the SAG sample have CaT*
values that are about 0.5~{\AA} lower than expected by theory, given
their ages and metallicities, signaling that some vital ingredient
might be missing in the models. For 5 Fornax dEs in our sample,
independently determined ages and metallicities can be found in the
literature. The age and metallicity measurements of \citet{rak01} are
based on narrowband photometry employing a modified Str\"omgren filter
system; \citet{hm94} use age and metallicity sensitive line-strengths
in the $\lambda\lambda 4000-5000$~{\AA} region. These studies find dEs
to be rather old ($\sim 10$~Gyr) and metal-poor ($-1.5<$ [Z/H]
$<-0.5$) stellar systems. Four out of these five dEs, due to the
near-constancy of CaT and CaT*, lie well above the predicted values,
by up to 2~{\AA} in the case of the most metal-poor dE. In fact,
FCC245 is the only galaxy in this subsample whose indices are in
agreement with its low metallicity content ([Z/H]$\approx -1.1$) and
old age ($t \approx 10$~Gyr). FCC245 has a low surface brightness
($\langle \mu_{\rm e} \rangle_B = 23.5$~mag/arcsec$^2$) and its P14
line is affected by bright sky emission lines (influencing PaT and
CaT* but not CaT), hence the somewhat larger errorbars. Although we do
not have ages and metallicities for the 11 other dEs in our sample,
one thing is clear;~9 have CaT*$\sim 7$~{\AA} which will place them on
the same sequence as the SAG ellipticals and the 4 high-CaT* dEs
presented in Fig.~\ref{fig_ssp}, irrespective of their ages or
metallicities. The locus of these dEs in the CaT*-Z plane is indicated
in Fig.~\ref{fig_ssp} by the upper shaded box.

Hence, the majority of the dEs in our sample follow the same CaT (or
CaT*) versus $\sigma$ or versus Z relations as bright ellipticals and
bulges of spirals. The observed near-constancy of CaT and CaT* over a
very wide range of metallicities ($-1.5<$ [Z/H] $<0.5$) is in flagrant
disagreement with the prediction that these indices should be
sensitive to metallicity. Three out of fifteen dEs in our sample have
CaT*$\sim 5-6$~{\AA}, in agreement with them being low-metallicity
galaxies (lower shaded box in Fig.~\ref{fig_ssp}). This makes our
observations even more puzzling;~apparently, at a given metallicity,
dEs with CaT*$\sim 5$~{\AA} and with CaT*$\sim 8$~{\AA} co-exist.

\section{Discussion and conclusions} \label{par_di}

In order to interpret their low CaT and CaT* in view of the current
models, SAG and CEN find that model uncertainties ($\sim
0.5$~{\AA}) are too small to provide a solution. They consider
variations of the IMF as a function of metallicity or velocity
dispersion, or Ca underabundance as a possible way out.

By varying the slope of a power-law IMF, CEN change the stellar
dwarf/giant ratio of the population. Since CaT* depends on the surface
gravity of stars, being low in dwarf stars and high in giant stars, a
high dwarf/giant ratio in the case of a high-Z SSP and a low
dwarf/giant ratio in the case of a low-Z SSP, could explain both the
low CaT* observed in bright Es and the high CaT* in dEs. However,
there is no sound physical basis for such variations of the slope of
the IMF. Theoretical derivations show that for stellar masses above
1~$M_\odot$, the IMF behaves as a Salpeter power-law and is almost
insensitive to the physical conditions within star-forming molecular
clouds \citep{pad02}. Only in the subsolar-mass regime does the
environment affect the stellar mass-distribution (e.g. low ambient
densities prevent the formation of very low-mass stars). In an old
stellar population the contribution of subsolar-mass stars becomes
more important. Using a piecewise power-law IMF with Salpeter slope
for masses above 0.6 $M_\odot$ and varying slope at lower masses, SAG
show that the steepening required to bring the model CaT* values in
accordance with those observed in bright Es, implies higher than
observed FeH~9916~{\AA} index values (FeH is strong in dwarf stars but
nearly abscent in giants) and too high stellar mass-to-light ratios
(see SAG and references therein). And if the IMF would vary with Z to
explain the high CaT* in dEs, why doesn't this effect manifest itself
also in low-Z, low-$\sigma$ systems such as globulars?

Moreover, dEs are not composed by SSPs but can have long and complex
star-formation histories (e.g. \citet{fer00}). If the IMF varies with
metallicity, it will also vary in time and it remains to be seen how
the use of variable IMFs in chemo-evolutionary models affects the
predicted properties of dEs. \citet{chiap00} explore a model for the
Galaxy using a metallicity-dependent power-law IMF which steepens as
the metallicity rises (slope $x = $log~Z$ + 4.05$), thus simulating
also a temperature dependency. They find that the model predictions
are in disagreement with important observational constraints and
conclude that such an IMF should be rejected. Using the theoretical
IMF proposed by \citet{pad97} to investigate the formation of
ellipticals, \citet{chiosi98} find that density is the leading
parameter for the variations of the IMF, instead of temperature.
This leads to dwarf/giant ratios that decrease with increasing
galactic mass, inverse to the trend invoked by CEN. Indeed, IMFs
with a low dwarf/giant ratio lead to numerous SN{\sc ii} explosions
that rapidly recycle heavy elements into the ISM and cause
$\alpha$-elements such as Mg, O and Ca to be overabundant with respect
to Fe. Also, such an IMF can raise the overall metallicity to above
solar.

Combining SSPs, SAG find that a rather artificial stellar mix has to
be invoked (consisting of a metal-rich stellar population with a small
contribution of a metal-poor one) in order to lower the predicted CaT*
to the one observed in bright Es and at the same time satisfy UV
constraints. Adding a (more logical) young metal-rich component to
explain the observed high H$\beta$ in ellipticals only aggravates the
problem, producing larger CaT* values. In dEs one could expect a small
high-metallicity stellar population on top of an old low-metallicity
population with the secondary star-burst triggered by
e.g. gravitational interactions with giant cluster members
\citep{mo96,con03b}. If the burst occured more than 2 Gyr ago, PaT
would be essentially unaffected. We explored this scenario by adding
young, high-metallicity SSPs to an old low-metallicity SSP and found
that any young, metal-rich SSP must contribute more than 50\% of the
total mass in order to raise CaT above 7.5~{\AA}. Then, the
young metal-rich population completely dominates the light, in
contradiction with the low metallicities inferred from optical
linestrengths.

Another suggestion to explain the low CaT* for Es is a true Ca
underabundance in giant galaxies. Although the optical Ca4227 Lick
index is also reported to be weaker than model predictions
(e.g. \citet{vaz97}), such an underabundance is hard to combine with
the observed $\alpha$-element overabundance for giant Es. Also, when
trying to reproduce the Ca underabundance with chemo-evolutionary
models, extremely short formation timescales ($\sim 10^7$ yr),
extremely flat IMFs ($x \sim 0.3$) \citep{mo00} or
metallicity-dependent supernova yields \citep{wo98} are required. On
the other hand, although the sensitivity of CaT* to the Ca abundance
is not yet fully understood, in this scenario the high CaT* in dEs
would imply a Ca overabundance at low metallicities (see \citet{tom03}
for a detailed discussion).

To conclude, we show that 12 out of a sample of 15 dEs follow the same
sequence as bright Es and bulges of spirals in CaT or CaT* versus
$\sigma$ diagrams whereas 3 out of 15 have significantly lower CaT*.
This could be the first direct evidence for the existence of two
distinct dE populations that are morphologically and kinematically
indistinguishable but have different chemical compositions. A possible
explanation for this dichotomy could be that dEs evolve to their
present-day state along different evolutionary paths. Two principal
evolutionary sequences are galaxy harassment
\citep{mo96,con03a} and SN-driven galactic winds \citep{ds86,ay87},
resulting in morphologically and kinematically similar stellar systems
but with distinct chemical properties. The kinematics of the full
sample will be presented in a forthcoming paper.

Variations of the IMF slope could explain the near-constancy of CaT*
over a wide range of metallicities. However, the proposed drastic
variations violate other observational
constraints and yield contradicting results when incorporated in
full-fledged chemo-evolutionary models. On the whole, it is clear that
we are still far from understanding the behavior of the \CaII triplet
in the integrated light of stellar systems with complex star-formation
histories. Only with the aid of detailed chemo-dynamical models can
one hope to explore the effects of realistic IMF variations and of
complex star-formation histories on the relative elemental abundances,
particularly of Ca and other $\alpha$-elements, and to check how
different formation scenarios reflect in the absorption-line indices
of dEs.

\acknowledgments 

We thank the anonymous referee for usefull comments.
% D.~M. acknowledges financial support of the Special Research Fund (BOF
% grant 011D6501).
W.~W.~Z. acknowledges the support of the Austrian Science Fund
(project P14783) and of the Bundesministerium f\"ur Bildung,
Wissenschaft und Kultur.
S.~D.~R. acknowledges financial support of the Belgian Fund for
Scientific Research.

\newpage
% figure 1

\newpage
% figure 2

\newpage
% figure 3

\end{document}